\documentclass[lettersize,journal]{IEEEtran}
\usepackage{amsmath,amsfonts}
\usepackage{algorithmic}
\usepackage{algorithm}
\usepackage{array}
\usepackage[caption=false,font=normalsize,labelfont=sf,textfont=sf]{subfig}
\usepackage{textcomp}
\usepackage{stfloats}
\usepackage{url}
\usepackage{verbatim}
\usepackage{graphicx}
\usepackage{cite}

\usepackage{soul}

\usepackage{amsthm}
\theoremstyle{definition}
\newtheorem{assumption}{Assumption}

\theoremstyle{definition}
\newtheorem{defn}{Definition}

\begin{document}
\title{Evaluating and Managing Tokenomics for Non-Fungible Tokens in Game-Based Blockchain Networks}

\author{Hyoungsung Kim, Hyun-Sik Kim, Yong-Suk Park
\thanks{Manuscript received}
\thanks{This research was supported by Culture, Sports and Tourism R\&D Program through the Korea Creative Content Agency grant funded by the Ministry of Culture, Sports and Tourism in 2022 (Project Name: Open Metaverse Asset Platform for Digital Copyrights Management, Project Number: R2022020034, Contribution Rate: 100\% )}
\thanks{Hyoungsung Kim is researcher of Contents Convergence Research Center, Korea Electronics Technology Institute (KETI), Seoul, Korea (e-mail: hyoungsung@keti.re.kr)}
\thanks{Hyun-Sik Kim is chief researcher of Contents Convergence Research Center, KETI, Seoul, Korea (e-mail: hskim@keti.re.kr)}
\thanks{Young-Suk Park is principal researcher of Contents Convergence Research Center, KETI, Seoul, Korea (e-mail: yspark@keti.re.kr)}
}

\markboth{Journal of \LaTeX\ Class Files,~Vol.~14, No.~8, August~2021}%
{Shell \MakeLowercase{\textit{et al.}}: A Sample Article Using IEEEtran.cls for IEEE Journals}

\IEEEpubid{0000--0000/00\$00.00~\copyright~2021 IEEE}

\maketitle

\begin{abstract}
Non-fungible tokens (NFTs) are becoming increasingly popular in Play-to-Earn (P2E) Web3 applications as a means of incentivizing user engagement. In Web3, users with NFTs ownership are entitled to monetize them. However, due to lack of objective NFT valuation, which makes NFT value determination challenging, P2E applications ecosystems have experienced inflation. In this paper, we propose a method that enables NFT inflation value management in P2E applications. Our method leverages the contribution-rewards model proposed by Curve Finance and the automated market maker (AMM) of decentralized exchanges. In decentralized systems, P2E Web3 applications inclusive, not all participants contribute in good faith. Therefore, rewards are provided to incentivize contribution. Our mechanism proves that burning NFTs, indicating the permanent removal of NFTs, contributes to managing inflation by reducing the number of NFTs in circulation. As a reward for this contribution, our method mints a compensation (CP) token as an ERC-20 token, which can be exchanged for NFTs once enough tokens have been accumulated. To further increase the value of the CP token, we suggest using governance tokens and CP tokens to create liquidity pools for AMM. The value of the governance token is determined by the market, and the CP token derives its value from the governance token in AMM. The CP token can determine its worth based on the market value of the governance token. Additionally, since CP tokens are used for exchanging NFTs, the value of the NFT is ultimately determined by the value of the CP token. To further illustrate our concept, we show how to adjust burning rewards based on factors such as the probability of upgrading NFTs' rarity or the current swap ratio of governance and CP tokens in AMM. Our mechanism provides a basis for evaluating NFTs and controlling their inflation which can serve as a fundamental business model for diverse P2E applications. In addition, our method allows P2E application governance to manage the inflation of NFTs in their ecosystem by evaluating NFTs.
\end{abstract}


\section{Introduction}
\IEEEPARstart{B}{lockchain} is a decentralized database that is distributed among the nodes of a peer-to-peer network [1]. To ensure trust among the peer nodes, a consensus algorithm is used to verify messages sent by other peers \cite{Eccpow}. Although the term ``consensus algorithm" has been used in distributed systems \cite{Paxos}, its meaning is different in the context of the decentralized system that blockchain exploits. In distributed systems, consensus algorithms mainly handle consistency, availability, or partition tolerance \cite{CAP Theorem} to resolve communication or network failures of peers. However, in blockchain systems, the consensus algorithm validates and links a transaction block that other peers send. These validations require continuous resource contributions from peers, such as computing power, electricity usage, hardware costs \cite{Bitcoin, Ethereum} or monetary staking \cite{Ethereum 2.0-1, Ethereum 2.0-2}. However, not all peers contribute in good faith, and the BitTorrent protocol \cite{BitTorrent} has empirically demonstrated the limitations of a network maintained solely by peers' good faith \cite{BitTorrent White paper}. Therefore, consensus algorithms in blockchain systems provide rewards to incentivize peers' contributions \cite{Blockchain and Cryptocurrencies}. For example, Bitcoin's consensus algorithm, Proof-of-Work, rewards contributors with Bitcoin.

\IEEEpubidadjcol

Applications of blockchain, known as decentralized apps (Dapps), also exploit contribution-reward models to build ecosystems for their services. Contributions may come in any form as long as they help a Dapp. Curve Finance \cite{Curve Finance}, a financial branch of Dapp, is a prime example of Decentralized Finance (DeFi) that runs a decentralized exchange (DEX), facilitating traders to swap their assets for other assets. Curve Finance's service uses various forms of contribution and rewards. Contributions may include not only cryptocurrencies but also forms of information that help to provide continuous service. For instance, users called liquidity providers to supply their assets as liquidity to Curve Finance, and in return, Curve Finance gives the liquidity providers a portion of the swap fees earned by the liquidity and liquidity pool tokens (LP tokens). LP tokens represent a liquidity provider's share of the pool and can be redeemed for the original assets at any time. However, such nonbinding liquidity supplying and claiming reduce the stability of Curve Finance's service. It can result in losing a lot of liquidity in a short period of time, similar to a bank run. To counter this, Curve Finance has adopted the practice of locking (i.e., preventing withdrawal) LP tokens for the long term in exchange for Decentralized Autonomous Organization (DAO) governance token \cite{Web3-DAO, Maker DAO, Smart contract DAO}. A DAO is a decentralized organization that operates using smart contracts on a blockchain and allows token holders to have a say in the decision-making process. Governance tokens are used for voting in a DAO, enabling token holders to vote for their profit or receive governance rewards. By locking LP tokens, Curve Finance can stabilize liquidity flows and securely plan its services. We provide more details on the contribution-reward model that Curve Finance is using in Section \ref{Section2: Background}.

DeFi has become one of the most successful Dapp services and has attracted cryptocurrency funding \cite{Perpetual contract NFT}, such as coins or standardized tokens. Coins are rewarded through consensus algorithms, while standardized tokens are tradeable assets minted by smart contracts that follow Ethereum's standard token interfaces, such as ERC-20 \cite{ERC-20} and ERC-721 \cite{ERC-721}. ERC-20 token interfaces are designed to hold value, while ERC-721 token interfaces, also known as Non-Fungible Tokens (NFTs), are designed to represent ownership rights. Therefore, users can use ERC-20 tokens to represent the assets' worth and ERC-721 tokens to carry digital deeds. The characteristic of ERC-20 token interfaces makes them widely used in DeFi, while the characteristic of ERC-721 does not \cite{Perpetual contract NFT}. DeFi services can assess the value of ERC-20 tokens because they hold the value of associated assets. For example, if an ERC-20 token is pegged to the U.S. dollar, its value is fixed at one U.S. dollar per token. However, the value of ERC-721 tokens relies on subjective evaluation or price history, making their value assessment difficult \cite{Perpetual contract NFT, NFT valuation, NFT pricing}. ERC-721 tokens are primarily used for digital collections (e.g., CryptoKitties) or representation of rights (e.g., Uniswap v3 \cite{Uniswap v3}), due to the difficulty in valuation. Perpetual contract NFTs \cite{Perpetual contract NFT} have been suggested as a method to assign value ERC-721 tokens using perpetual contracts, but they are limited to DeFi services. Dapp services such as Axie Infinity \cite{Axie infinity} and STEPN \cite{STEPN} have demonstrated that ERC-721 tokens can also be utilized in Play-to-Earn (P2E) applications, where users can earn money by selling in-application goods via blockchain. However, the difficulty of valuation makes them challenging to handle in terms of NFT inflation. Therefore, further research and development are needed to create better methods for valuing and utilizing ERC-721 tokens in P2E applications.

In this paper, we propose a method for evaluating and handling inflation for NFTs in incentivized P2E Web3 applications, inspired by the Automated Market Maker (AMM) of Uniswap \cite{Uniswap V2} and the contribution-rewards model of Curve Finance \cite{Curve Finance}. Web2 refers to traditional web services, while Web3 refers to emerging platforms \cite{DeSci Web3} that utilize blockchains, cryptocurrencies, and NFTs. One significant difference between Web2 and Web3 is ownership \cite{Introduction to Web3}. In Web3, NFTs provide users with ownership rights to their digital goods. Several NFT interface standards inherit the original ERC-721 standard, such as ERC-1155\cite{ERC-1155}, which is also known as the multi-token standard. In this paper, we use the term NFT to refer to both ERC-721 and ERC-1155 standards. While ERC-721 represents a single unique token with a unique identification (ID) number, ERC-1155 allows for multiple tokens to be associated with a single ID number. However, if a user only mints one NFT with ERC-1155, it functions similarly to an ERC-721 NFT. The decision of which NFT standard to use depends on the P2E Web3 application business model (BM).

AMM algorithms, proposed by Uniswap, are commonly used to create liquidity pools for ERC-20 tokens in DEXs. Since many DEXs utilize AMM algorithms for ERC-20 token pools, our work can leverage previous studies in this area. While widely used in various DEX services, including Curve Finance, AMM has not been applied to ERC-721 tokens because of the difficulty of valuation. While Sudoswap\cite{What is Sudoswap} attempted to adopt AMM for NFTs, their focus was solely on decentralized NFT trading. Furthermore, liquidity providers using AMM algorithms typically require cryptocurrency pairs for initial provision, which poses limitations for pools consisting of different standard tokens, such as ERC-721 and ERC-20 token pairs, in terms of liquidity collection. In this work, we propose an approach for creating a pool with ERC-20 token pairs for NFTs, enabling the collection of liquidity. By using the pool, we present a method for valuating NFT and controlling its inflation in P2E Web3 applications by leveraging AMM and the contribution-rewards model of Curve Finance.

The paper is organized as follows: Section \ref{Section2: Background} provides background information on AMM and the contribution-rewards model of Curve Finance to help readers understand the basis of our work. In Section \ref{Section3: Leveraging}, we present our proposed method for leveraging the AMM and contribution-rewards model of Curve Finance to value and manage the inflation of NFTs in incentivized P2E Web3 applications. Finally, Section \ref{Section4: Conclusion} summarizes our work and provides conclusions.

\section{Background}
\label{Section2: Background}
AMM algorithms, proposed by Uniswap, are widely used in various DEXs to swap ERC-20 tokens automatically. Curve Finance, one of the more popular DEXs, also uses specialized AMM for exchanging stablecoins, ERC-20 tokens pegged to fiat currencies. Additionally, Curve Finance employs a contribution-rewards model underlying decentralization for their governance. This model incentivizes liquidity providers by rewarding them with governance tokens and controls token inflation through DAO voting. Our work proposes combining NFTs and governance tokens on AMM to evaluate NFTs and control their inflation. The AMM algorithm has undergone significant improvements and has been adopted by various DEXs in recent years. Moreover, Curve Finance has been continually upgrading its protocol through innovative research. Therefore, in this section, we aim to provide a detailed explanation of the underlying principles of these concepts that have served as a foundation for our work.

\subsection{Automated Market Maker}
\label{Section2.1: AMM}

\begin{figure}[!t]
\centering
\includegraphics[width=1\linewidth]{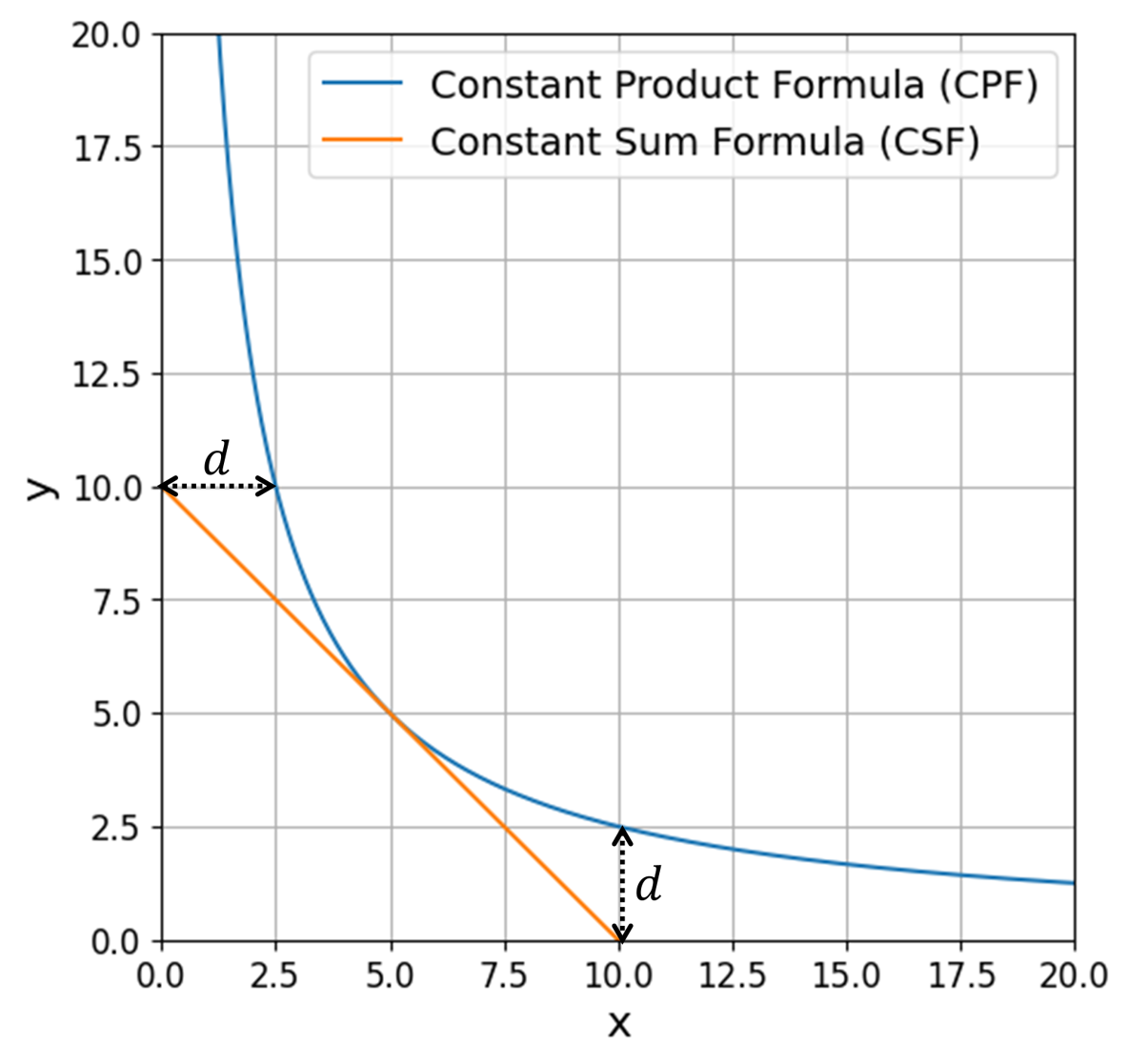}
\caption{Comparison of two Automated Market Makers (AMMs): Constant Product Formula (CPF) and Constant Sum Formula (CSF), both with an ideal swap ratio of 1:1 for assets X and Y. Initial liquidity is set to $x=5$ and $y=5$. Research has been conducted to improve the CPF, with the goal of decreasing the deviation distance $d$ between CPF and CSF.}
\label{fig:AMM}
\end{figure}

In centralized exchanges (CEXs), cryptocurrencies are traded based on matching prices (i.e., bid and ask) provided by intermediaries. However, in decentralized exchanges (DEXs) based on Automated Market Makers (AMMs), traders swap their assets with staked assets in a liquidity pool. Liquidity is provided by decentralized liquidity providers who earn a fee in exchange. AMM algorithms adjust the swap ratio of the liquidity pool to provide automated swapping to traders. Liquidity pools usually consist of pairs of assets provided by liquidity providers. For example, (X, Y) liquidity pool holds ERC-20 tokens X and Y. Traders can swap X with Y at a determined swap ratio and vice versa. The AMM algorithm automatically adjusts the swap ratio to minimize slippage, which is the loss incurred when the expected swap ratio of traders differs from the actual swap ratio provided by the liquidity pool. Slippage is determined by how efficiently an AMM exploits the liquidity in a pool, and the lower the slippage, the better the AMM at optimizing trades for users. Hence, one of the primary goals of AMM is to reduce slippage. The most well-known AMM is the constant product formula (CPF) proposed by Uniswap, which maintains the product of two reserved assets at a constant value:

\begin{equation}
\label{equation:CPF}
 x \cdot y = k
 \end{equation}

\noindent
 where $x$ is the quantity of asset X, $y$ is the quantity of asset Y, and $k$ is a constant. Many DEX projects, such as Curve Finance, improved CPF AMMs to reduce slippage and adapt them to their BMs. In CPF, the swap ratio between assets X and Y is maintained by ensuring that their product of reserves remains constant. For instance, if the initial liquidity of X and Y is five each, the product of their reserves is 25. When a trader wants to purchase one unit of X, they need to pay an amount of Y that preserves the constant value of 25. Thus, the trader should pay $y^\prime = 1.25$ by solving $(5-1)\cdot(5+y^\prime)=25$.
 
 As the amount of liquidity in the pool increases, the amount of Y, that users should pay for one unit of X, decreases. For example, if X and Y both have an initial liquidity of 500, a user can buy one unit of X with only 1.002 Y. The difference in price resulting from the amount of liquidity in the pool is what is referred to as slippage. In the theoretical case of infinite liquidity in a liquidity pool, DEXs can use the Constant Sum Formula (CSF) to achieve zero slippage: $x + y = k$. Fig. \ref{fig:AMM} shows the difference between CPF and CSF. As the amount of liquidity in the pool increases, the distance $d$ between CPF and CSF decreases. However, in reality, liquidity in a liquidity pool is limited, and large trades can cause a significant price impact, resulting in depleting liquidity for subsequent trades if DEXs use the CSF as the AMM to adjust the swap ratio of a liquidity pool. Therefore, DEXs have focused on developing AMM protocols based on the CPF to reduce slippage.

The improved CPF underlying Curve Finance is specifically designed for stablecoins, which are cryptocurrencies that are pegged to the value of a fiat currency such as the US dollar. Due to stablecoins are designed to maintain a stable value, their prices tend to fluctuate less than other cryptocurrencies, making them ideal for use in financial applications such as trading and lending. To reduce slippage and improve the efficiency of trades, the AMM of Curve Finance allocates more liquidity to a range of swap ratios where swaps are most commonly made rather than allocating liquidity evenly across all possible swap ratios. This approach allows the AMM to function more like a CSF in that range while still maintaining the benefits of the CPF for other swap ratios. By optimizing liquidity allocation, Curve Finance has been able to provide traders with more favorable prices and reduce the impact of large trades on liquidity.

\subsection{Contribution-rewards model of Curve Finance}
\label{Section2.2: Contribution-rewards model}

\begin{figure*}[!t]
\includegraphics[width=0.9\textwidth]{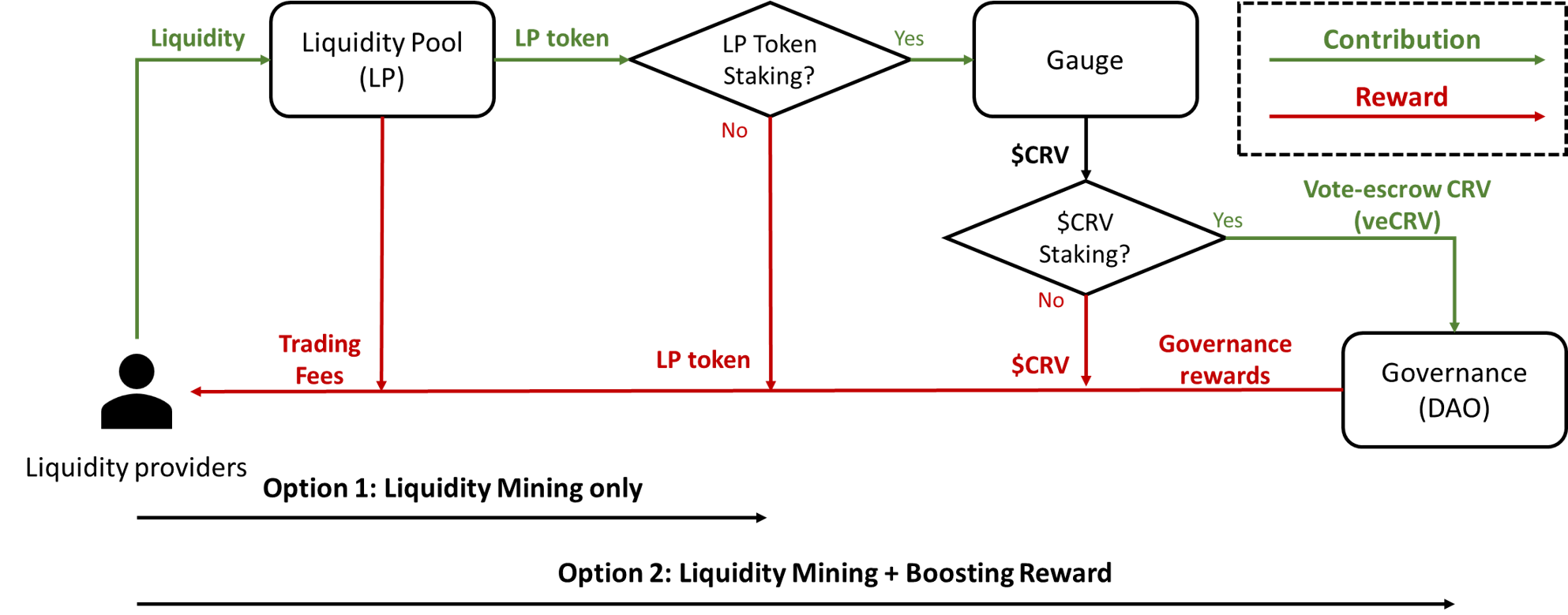}
\centering
\hfil
\caption{In Curve Finance, liquidity providers have two options for contribution and rewards. For the short term, they can choose \textbf{Option 1}, which allows them to claim their liquidity quickly. For the long term, they can choose \textbf{Option 2}, which includes boosted rewards. With this option, liquidity providers stake their liquidity for a longer period and cannot claim it until the expiration date, but they receive higher rewards.}
\label{Fig:Curve Finance Contribution-rewards model overview}
\end{figure*}

Decentralized systems, including Dapps, require appropriate contribution-rewards models to incentivize users to contribute. These systems do not have a central authority forcing incentives, so without a proper contribution-rewards model, they may cease to function effectively. One example of such a decentralized system is Filecoin\cite{Filecoin}, a decentralized storage network based on InterPlanetary File System (IPFS) \cite{IPFS, IPFS design}. IPFS is a distributed peer-to-peer file system that connects all computing devices with the same system of files. To incentivize users to contribute their storage space for IPFS, Filecoin uses a contribution-rewards model where FIL, the name of Filecoin reward, is given to users as a reward for their contributions. This approach has enabled Filecoin to provide a reliable and decentralized storage service since its launch in 2017.

Curve Finance has designed a contribution-rewards model for its Dapp, as shown in Fig. \ref{Fig:Curve Finance Contribution-rewards model overview}. When a {\fontfamily{Calibri}\selectfont Liquidity providers} deposit their cryptocurrency as liquidity to a {\fontfamily{Calibri}\selectfont Liquidity Pool}, they become a market maker for Curve Finance's pool. {\fontfamily{Calibri}\selectfont Liquidity providers} receive two rewards for a market-making contribution: a portion of the accumulated trading fees in proportion to their liquidity shares and the liquidity pool token (LP token), which represents the liquidity deposited by the provider. Each liquidity pool has its unique LP token. {\fontfamily{Calibri}\selectfont Liquidity providers} have two options after depositing liquidity:

\begin{itemize}
\item{{\fontfamily{Calibri}\selectfont \bf{Option 1}} provides a portion of the trading fees proportional to their shares, and {\fontfamily{Calibri}\selectfont Liquidity providers} hold the LP token to claim their deposit whenever they want.}
\item{{\fontfamily{Calibri}\selectfont \bf{Option 2}} offers multiple contribution-reward pairs. {\fontfamily{Calibri}\selectfont Liquidity providers} convert LP tokens to \$CRV tokens by staking LP tokens for a set period. \$CRV is the governance token of Curve Finance.}
\end{itemize}

\noindent
\textbf{Options 1} and \textbf{2} in Curve Finance can be compared to a savings account and a fixed deposit in traditional banks, respectively. \textbf{Option 1} is like a savings account, where liquidity providers can earn small rewards without committing to a fixed tenure. This makes it an attractive option for those who want short-term yields. On the other hand, \textbf{Option 2} is like a fixed deposit, where liquidity providers must stake their funds for a predetermined amount of time. This is comparable to a fixed deposit in a bank, where customers must commit to a fixed tenure to earn higher interest rates. In both cases, the liquidity providers are contributing to the stability and growth of the platform, like how bank customers are contributing to the stability and growth of the bank by depositing their funds.

In \textbf{Option 2}, the liquidity provided by the users is measured by the gauge system. In Curve Finance, the gauge is a mechanism used to measure liquidity movement and distribute \$CRV tokens for each LP token. This helps control inflation in Curve Finance. Liquidity providers may use the \$CRV rewarded for DeFi composability \cite{Perpetual contract NFT} in other Dapps or continue contributing by staking \$CRV. The gauge system can measure the inflation of \$CRV by measuring how many tokens are staked. Staking \$CRV gives liquidity providers vote-escrow CRV (veCRV), which translates to voting power in DAO Governance. The longer the staking period, the more veCRV earned, resulting in increased voting power. This motivates users with lower veCRV to increase their voting power by staking \$CRV for a longer period.

\section{Leveraging AMM and Contribution-rewards model for P2E Web3 Applications}
\label{Section3: Leveraging}
Before we present our proposed method for valuating and controlling inflation of NFTs in incentivized P2E Web3 applications, we outline the following definitions and assumptions that guide our work:

\begin{defn}[Burning NFT]
    ``Burning NFT" is a term used to describe the process of sending an NFT to a designated address, usually the ``zero address" (an address consisting of 20 bytes of zeros with the prefix ``0x"), with the intention of permanently removing it from circulation. It is currently understood that the private key of the zero address is unknown to anyone.
\end{defn}

\begin{defn}[Trustless NFT counters]
    The quantity of NFTs in circulation can be publicly monitored by all market participants through the blockchain, without the need for trusted intermediaries.
\end{defn}

After defining key terms used in Dapps with \textbf{Definitions 1} and \textbf{2}, we introduce a set of assumptions that help establish conditions for the analysis of Dapps. Specifically, our assumptions include the scarcity of NFTs increasing their value, rational governance of DAOs, and an initial value for NFTs, as stated in \textbf{Assumptions 1} to \textbf{3}.

\begin{assumption}[Scarcity of NFT increases value]
    The value of an NFT increases when its scarcity is greater than that of other NFTs.
\end{assumption}

\begin{assumption}[Rational DAO governance]
    In a DAO governance system, participants are assumed to be rational actors who will not intentionally make decisions that harm the governance or its stakeholders.
\end{assumption}

\begin{assumption}[Initial NFT value]
    When a new NFT is minted, the Rational DAO sets an initial value for the NFT that is based on market factors, including rarity, demand, and other relevant considerations. This initial value is set at a reasonable level, in line with market norms, and reflective of the unique qualities of the NFT.
\end{assumption}

We propose a method that combines AMM and the contribution-rewards model for NFTs in P2E Web3 applications, such as Axie Infinity \cite{Axie infinity} and STEPN \cite{STEPN}, which typically use probability to control rarity and inflation. Users can obtain rarer NFTs by enchanting their NFTs or by drawing from a randomized reward system. In order to control inflation, users may need to burn NFTs or in-game goods for the enchanting or drawing. Our approach considers burning NFTs as a contribution that helps to control inflation, and by leveraging this contribution, we provide an efficient and effective way to manage NFT value and control inflation in P2E applications.
 
\subsection{Leveraging AMM and contribution-rewards model}
Burning NFTs reduces the number of NFTs in circulation, thereby controlling their scarcity and contributing to inflation control. Therefore, users who burn their NFTs should receive be rewarded. To achieve this, we propose adopting the LP token concept used in Curve Finance. Each liquidity pool has its own LP token, which can be exchanged for a governance token by staking it to the Gauge. If governance allocates a token that functions similarly to the LP token for each NFT group, where grouping may be based on rarity or predefined category, a portion of the token can be provided as a reward for burning NFTs. This approach incentivizes users to participate in NFT inflation control by providing them with a tangible reward for their efforts.

\subsubsection{Integrating AMM for NFT valuation}
\label{subsubsection: Integrating AMM for NFT valuation}
We propose using a compensation (CP) token as an ERC-20 token to represent the value of NFTs. Each NFT group will have its own CP token, and governance will reserve CP tokens as rewards for burning and as liquidity for an AMM. When users burn NFTs, they will receive CP tokens as burning rewards. Users can also purchase NFTs by paying CP tokens that reflect the value of the NFT, in much the same way as how they would claim liquidity using LP tokens. Unlike Curve Finance, where LP tokens are burned to claim liquidity, our approach replenishes burning rewards with CP tokens. This helps prevent the depletion of reserved CP tokens given as rewards for NFT burning. CP tokens allocated for the AMM's liquidity will be paired with a governance token. The value of CP tokens will be determined by the swap ratio between CP tokens and governance tokens. For example, if one governance token is worth \$10 in centralized exchanges, the swap ratio of the governance token and CP token is 1:10 in an ideal AMM. Since there is no slippage, the value of the CP token will be pegged to 0.1 governance token by the AMM. With this system, users can purchase an NFT with 10 CP tokens, and the value of the NFT will be close to \$10. This approach provides a transparent and efficient way to manage NFT value in P2E applications.

To create a liquidity pool, we propose integrating the CP token and governance token of the P2E application into the AMM equation (\ref{equation:CPF}). In order to reserve a portion of CP tokens for burning rewards, only a specific amount of CP tokens should be provided as liquidity. By \textbf{Assumptions 2} and \textbf{3} the rational DAO will define a portion of CP tokens to be included in the liquidity pool. The resulting equation for the initial CPF AMM for NFT valuation is as follows:

\begin{equation}
\label{equation:CP token AMM}
(q \cdot cp) \cdot g = k
\end{equation}

\begin{figure}[!t]
\centering
\includegraphics[width=1\linewidth]{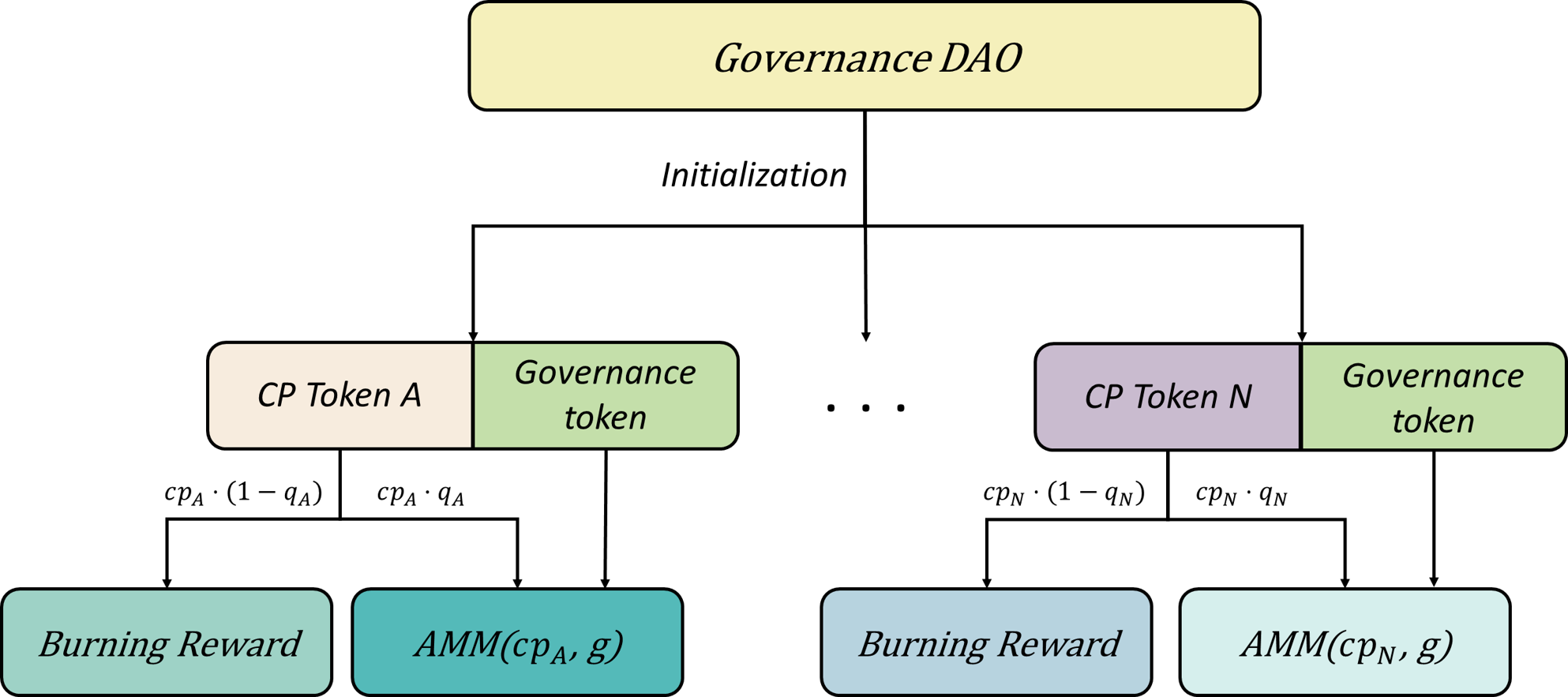}
\caption{{\fontfamily{Calibri}\selectfont $Governance$ $DAO$} initializes CP tokens and allocates governance tokens for each NFT group. The amount of CP tokens allocated for {\fontfamily{Calibri}\selectfont $Burning$ $Reward$} or {\fontfamily{Calibri}\selectfont $AMM$} depends on the value of $q$.}
\label{fig:Initializing_pool}
\end{figure}

\noindent
The variable $cp$ represents the total value of existing CP tokens, while $q \cdot cp$ denotes the fraction of the total CP tokens value allocated as liquidity, where $q$ is a value between 0 and 1. On the other hand, the value of governance token provided as liquidity is represented by the variable $g$. Fig. \ref{fig:Initializing_pool} illustrates the initialization of burning rewards and AMMs for each NFT group. The initialization of the {\fontfamily{Calibri}\selectfont $Governance$ $DAO$} includes initializing the CP token, allocating governance tokens for each NFT group, deciding on the value of $q$ for allocating CP tokens to the {\fontfamily{Calibri}\selectfont $Burning$ $Reward$}, and creating an {\fontfamily{Calibri}\selectfont $AMM$}. Each NFT group has its own CP token, such as {\fontfamily{Calibri}\selectfont $CP$ $token$ $A$}, but they all use the same governance token.

Let’s assume an NFT group where CP tokens are allocated. There are $x$ NFTs in the group with the same rarity level. For the sake of intuitive understanding, assume that there is no slippage in the AMM. Therefore, NFT holders can receive $(1-q) \cdot cp / x$ CP tokens as a reward for burning their NFT. If the holder then brings those CP tokens to the AMM to convert them to governance tokens, (\ref{equation:CP token AMM}) will be given to the holder as:

\begin{equation}
\label{equation:valuation}
(q \cdot cp + \frac{(1-q) \cdot cp}{x}) \cdot (g - g \prime) = k
\end{equation}

\noindent
where $g \prime$ represents the value of the governance token that the CP token holder obtains through the AMM. Equation (\ref{equation:valuation}) can be rearranged to calculate the $g \prime$ as follows:

\begin{equation}
\label{equation:value of g prime}
    g \prime = g - \frac{k}{q \cdot cp + C}
\end{equation}

\noindent
The constant term $C$ represents $(1-q) \cdot cp / x$. Equation (\ref{equation:value of g prime}) presents the value of the governance token that can be obtained from the AMM. The value of the governance token is determined by the market, and the CP token can leverage this value to determine its own worth, as the value of the CP token is ultimately derived from the value of the governance token. Finally, the value of the NFT is determined by the value of the CP token because CP tokens are used to exchange with the NFT. In reality, factors such as slippage and transaction fees need to be considered, but this example illustrates a basic mechanism of how the CPF AMM can be used to assign value to NFTs.

\subsubsection{Integrating contribution-rewards model for controlling inflation}
\begin{figure*}[!t]
\includegraphics[width=0.9\textwidth]{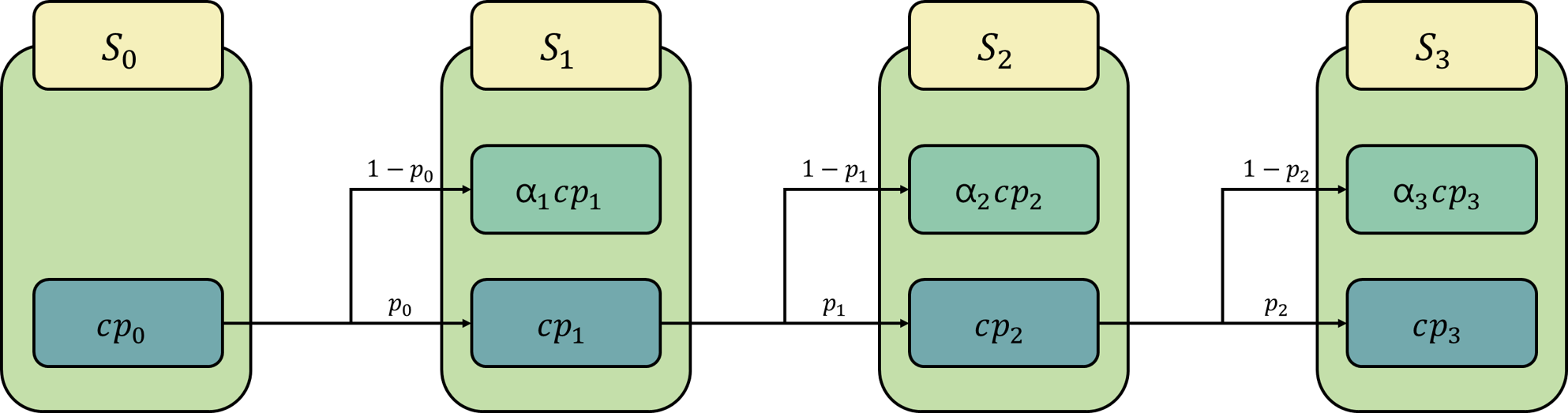}
\centering
\hfil
\caption{The value of an NFT at rarity state $S_i$ is denoted by $cp_i$, where $S_0$ is the initial rarity state and $cp_0$ is the initial value. With probability $p_0$, an NFT representing a value of $cp_0$ will be upgraded to a new rarity state $S_1$, with the corresponding value $cp_1$. If the NFT holder fails to upgrade, with probability $1-p_0$, they receive a burning reward in the form of $\alpha_1 cp_1$ tokens, where $\alpha_i$ is a decaying factor used to control inflation.}
\label{fig:Reward state diagram}
\end{figure*}

In our demonstration of integrating AMM for NFT valuation, we assumed that all NFTs sharing a liquidity pool had equal rarity and received the same amount of burning rewards. However, in reality, P2E applications present NFTs having different values and incentivize users to obtain or upgrade to rarer NFTs. These applications typically control the inflation of NFTs in the upgrade process by using probability. If users fail to upgrade, their NFTs are burnt or damaged, consequently decreasing the value of NFTs in circulation. Therefore, burning rarer NFTs should yield more significant rewards to incentivize upgrading NFTs to help control inflation. If rare NFTs gave fewer rewards when burned, NFT holders would be discouraged from doing so, and the number of rare NFTs would accumulate, causing their value to decrease. Hence, we propose a way to set the rewards of a contribution-rewards model to control inflation in NFTs with changing rarity, allowing for greater flexibility in NFT valuations, particularly in P2E Web3 applications. 

Fig. \ref{fig:Reward state diagram} shows a rarity state diagram for rewards. The rarity state of the $i$-th upgrade is denoted by $S_i$, where $i$ is a non-negative integer, and $S_0$ represents the initial rarity state of an NFT. With a probability of $p_i$, users can upgrade rarity state $S_i$ to $S_{i+1}$. If users successfully upgrade, $cp_i$, representing the value of the NFT expressed as CP tokens at the state $S_{i}$, is updated to $cp_{i+1}$ at state $S_{i+1}$. On the other hand, with a probability of $1 - p_i$, users fail to upgrade, lose the NFT, and receive $\alpha_{i+1} cp_{i+1}$ as burning rewards. The $\alpha_{i+1}$ is the decaying factor at state $S_{i+1}$ used to control inflation. To prevent inflation, we need to ensure that the expected value of the next state, which is calculated as $E[S_{i+1}] = p_i cp_{i+1} + (1-p_i) \alpha_{i+1} cp_{i+1}$, is less than or equal to the current state value, represented as

\begin{equation}
\label{equation:expectation condition}
    E[S_{i+1}] \le cp{_i}    
\end{equation}

\noindent
If the expected value exceeds $cp{_i}$, then the total value in circulation increases, and burning NFTs becomes ineffective in controlling inflation. The decaying factor $\alpha_{i+1}$ of $E[S_{i+1}]$ should be adjusted by the P2E Dapp BMs. We can further illustrate our concept by defining $\alpha_{i+1}$ as the product of $p_{i}$ and $I$, where $p_i$ is the probability of transition from state $S_{i}$ to $S_{i+1}$, and $I$ is the inflation factor. The inflation factor $I$ represents the number of governance tokens obtained by swapping CP tokens in the AMM curve. It should be noted that the inflation factor $I$ is assumed to be updated according to a predefined schedule set by the DAO, depending on the current swap ratio of the governance token and CP token in AMM.

For example, if the ideal swap ratio of the governance token and CP token is 1:1, but the current swap ratio is 1:2, this indicates that there is an oversupply of burning rewards in circulation. As a result, $I$ is updated to $1/2$, and burning rewards are halved. Conversely, if the swap ratio is 2:1, in case of scarcity, burning rewards increase. Moreover, setting the decaying factor, $\alpha_{i+1}$, with $p_{i}$ has the additional benefit of compensating unlucky users. The reward is necessary to motivate users to continue participating in the process, even if they have been unsuccessful multiple times. For instance, if the probability of success is 0.1 and $I$ is one, then $\alpha_{i+1}$ is 0.1. This means that even if a user fails ten times in a row, they can still obtain a similar value of upgraded rarity by using burning rewards.


\subsection{Valuation of Upgraded NFTs}
Our proposed contribution-rewards model with burning rewards aims to control inflation by maintaining an appropriate valuation, as specified in (\ref{equation:expectation condition}). To successfully implement the contribution-rewards model with burning rewards, market users need to be able to recognize the reasonable value of NFTs, which is determined by probability. We propose that the value of an upgraded NFT from rarity state $S_i$ to $S_{i+1}$ can be expressed as the product of the enhanced rarity value factor $r_i$ and the current rarity value $cp_i$, as given by the equation: $cp_{i+1} = r_i cp_i$. Thus, (\ref{equation:expectation condition}) can be rearranged to obtain the maximum possible value of the upgraded NFT as follows:

\begin{equation}
    \label{eqaution:Updated expectation condition}
    p_i r_i cp_i + I (1-p_i) p_i r_i cp_i \le cp_i
\end{equation}

\noindent
By rearranging terms, we can calculate the ceiling for possible enhanced rarity value factor of the upgraded NFT:

\begin{equation}
    \label{eqaution:Possible rarity enhancement}
    r_i \le \frac{1}{p_i + I(1-p_i)p_i}
\end{equation}

\begin{figure}[!t]
\centering
\includegraphics[width=0.9\linewidth]{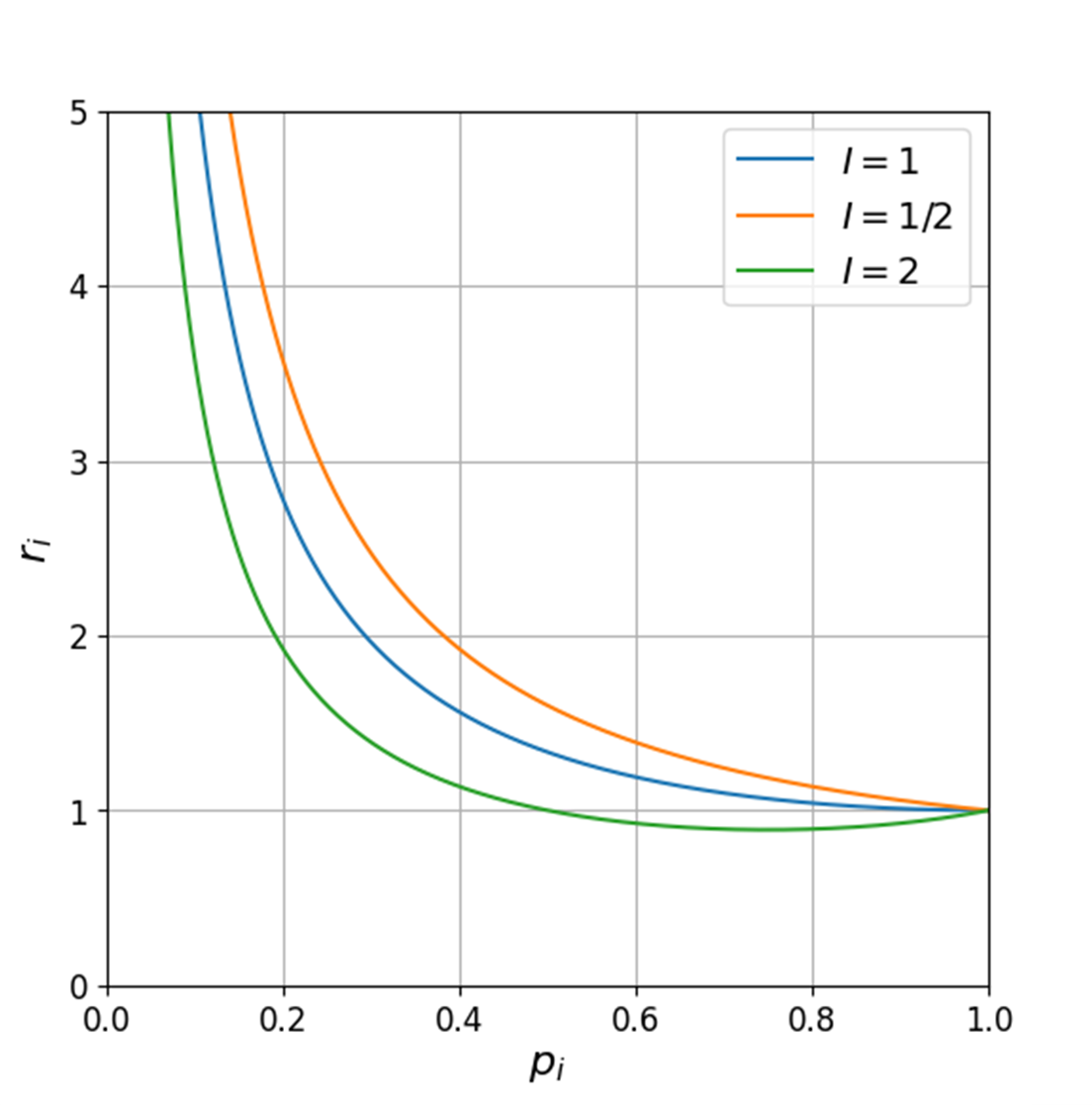}
\caption{The enhanced rarity factor can be determined for different values of $\{1/2, 1, 2\} \in I$. This enables market users to determine a reasonable value for upgraded NFTs based on the current value of $I$.}
\label{fig:Reward_valuation}
\end{figure}

\noindent
Fig. \ref{fig:Reward_valuation} displays the enhanced rarity value factor over $p_i$ for different values of $\{ 1/2, 1, 2\} \in I $, assuming an ideal swap ratio of 1:1. As the governance DAO adjusts $I$ from $1$ to $2$, indicating that the current swap ratio between governance token and CP token approaches 2:1, the enhanced rarity value factor decreases due to the burning reward $\alpha_{i+1}cp_{i+1}$ becoming twice as large, thereby relieving the scarcity of CP tokens.

The governance DAO determines the value of the initial state $S_0$ when they create an AMM pool. Hence, user can decide whether to buy or sell in a market by calculating the maximum value of $k$ times the value of the upgraded NFT, $k \ge 1$ as follows:

\begin{equation}
    \label{equation:Maximum value}
    cp_k = cp_0 \prod_{n=0}^{k-1} r_n
\end{equation}

\noindent
The profitability of a given valuation by (\ref{equation:Maximum value}) may attract arbitrageurs looking to exploit any discrepancies between the AMM and external markets, such as centralized exchanges (CEXs). For instance, if the swap ratio of governance token and CP token is not ideal, arbitrageurs can buy tokens, such as governance tokens, CP tokens, or CP token exchangeable NFTs, from the external market and swap them in the AMM. Once the AMM reaches the ideal swap ratio, they can swap back to increase the number of tokens to sell in the external market. Ultimately, such arbitrage activity and adjustments to the inflation factor can help guide the AMM toward the ideal swap ratio.

\subsection{Discussion}
The AMM and contribution-rewards model with burning rewards can enable the valuation of NFTs in P2E applications. To achieve this, it is important to tailor the probability and factor calculations to the specific BMs being used. In this regard, we present two considerations that should be taken into account when implementing the contribution-rewards model with burning rewards in P2E applications.

First, P2E applications should decide whether the total number of governance tokens is finite. To ensure fairness in governance, it is important to take into account the contributions of early participants, who may have contributed more than later participants. Gradually decreasing the number of rewarding governance tokens over time, as typically controlled by the governance DAO, is one way to address this issue in Dapps, such as DeFi. However, the finite number of governance tokens may limit the generation of liquidity pools for P2E applications. In P2E applications that leverage AMM, governance should create liquidity pools for AMM when new types of NFTs are launched in the application. If the supply of governance tokens is gradually decreasing, the number of tokens allocated as rewards for new types of NFTs will be less than for existing NFTs. This may lead to a situation where new types of NFTs are perceived as less valuable than existing NFTs. Some BMs may allow new types of NFTs that are less valuable, while others may require equal value regardless of time. Therefore, if the total number of governance tokens is finite, the NFT launching plan should carefully consider the number of governance tokens allocated to a liquidity pool to ensure that the value of new NFTs is properly reflected in the market.

On the other hand, if governance tokens are infinite, an additional burning method similar to that used by other decentralized systems or applications should be adopted to ensure fairness in governance. In the case of coins or tokens, where the value is the same regardless of the order of participation, Ethereum's method of minting Ether without limitation but burning it adaptively based on network congestion \cite{Ethereum London} can be used as a reference. In the case of governance tokens, a fair reward system is essential to incentivize early contributors and maintain their motivation to participate in governance. Therefore, it is crucial to ensure that those who contribute significantly earlier receive greater rewards. However, without an appropriate burning governance token method, the value of governance tokens may be subject to inflation, which can reduce the motivation of contributors.

Second, P2E applications need to determine the amount of CP tokens to reserve as burning rewards. Equation (\ref{equation:CP token AMM}) shows that a liquidity pool for AMM requires reserving $q \cdot cp$ tokens, leaving $(1-q) \cdot cp$ tokens as burning rewards. In addition, when the number of NFTs sharing AMM, $C$, is fixed, governance can use the probability mass function (PMF) $X(i)$, where $i\geq 1$, to determine the expected number of NFTs to burn at each rarity, which is $C\cdot X(i)$, where $X(i)$ is as follows:

\begin{equation}
    \label{failure distribution}
    X(i) = 
    \begin{cases}
1-p_0, & \text{if }i = 1 \\
(1-p_{i-1})\prod_{n=0}^{i-2}p_n , & \text{if } i \ge 2
\end{cases}    
\end{equation}

\noindent
Using (\ref{eqaution:Possible rarity enhancement}) and (\ref{equation:Maximum value}), governance can then calculate the maximum possible reward per NFT burning. Finally, by considering the expected number of NFT burning and the maximum possible reward per burning, governance can determine the appropriate amount of rewards to reserve. To ensure that the burning rewards remain fair and up-to-date, they can be adjusted regularly, e.g., every 24 hours, based on various BM factors. These factors could include the number of NFTs currently in circulation, the number of NFTs that are yet to be minted, or other relevant metrics.

\section{Conclusion}
\label{Section4: Conclusion}
In this paper, we present a decentralized NFT valuation method that leverages AMM in incentivized P2E Web3 applications. Our method uses governance tokens and CP tokens, given as rewards when NFTs are burnt, to allocate liquidity pools in AMM for NFT valuation. Additionally, our contribution-rewards model defines NFT burning as a valuable contribution that helps control inflation and rewards users for their contributions.

We also introduce a probabilistic mechanism for reserving CP tokens as burning rewards based on the probability of success, which incentivizes users to contribute by burning NFTs and helps prevent NFT inflation. Our proposed method can help P2E applications achieve a more fair and sustainable governance system, which can attract more users and increase the value of the ecosystem.

Future research will explore the effectiveness of this method in real-world P2E applications and evaluate its impact on the governance and economic aspects of these applications. Overall, our proposed method offers a promising approach to decentralized NFT valuation that can enhance the user experience and provide a more sustainable and equitable ecosystem for P2E applications in Web3.

\vfill

\end{document}